\documentclass[a4paper,fleqn,usenatbib]{mn2e}

\usepackage{newtxtext,newtxmath}

\usepackage[T1]{fontenc}
\usepackage{ae,aecompl}

\usepackage{graphicx}	
\usepackage{amsmath}	
\usepackage{amssymb}	
\usepackage{pdflscape}  
\usepackage{longtable}  

\def\braket#1{\left<#1\right>}
\defcitealias{yil14}{Paper~I}
\defcitealias{yil15}{Paper~II}
\defcitealias{yil16}{Paper~III}
\defcitealias{yilprep}{Paper~IV}

\title[Asteroseismic Investigation of Host Stars]{Asteroseismic Investigation of $20$ Planet and Planet-Candidate Host Stars}

\author[C. Kayhan, M. Y{\i}ld{\i}z and Z. \c{C}elik Orhan]{
C. Kayhan\thanks{E-mail: cenkkayhan@gmail.com},
M. Y{\i}ld{\i}z
and Z. \c{C}elik Orhan
\\
Department of Astronomy and Space Sciences, Science Faculty, Ege University, 35100 Bornova, {\.I}zmir, Turkey\\
}

\date{Accepted XXX. Received YYY; in original form ZZZ}

\pubyear{2019}

\begin{document}
\label{firstpage}
\pagerange{\pageref{firstpage}--\pageref{lastpage}}
\maketitle

\begin{abstract}
Planets and planet candidates are subjected to great investigation in recent years. In this study, we analyse 20 planet and planet-candidate host stars 
at different evolutionary phases. 
We construct stellar interior  models of 
the host stars with the {\small MESA} evolution code and obtain 
their fundamental parameters under influence of observational 
asteroseismic and non-asteroseismic 
constraints. Model mass range of the host stars is 0.74-1.55 $\rm M_{\sun}$.
The mean value of the so-called large separation between oscillation frequencies and its variation about the minima show
the diagnostic potential of asteroseismic properties.
Comparison of variations of model and observed large separations versus the oscillation frequencies 
leads to 
inference of fundamental parameters of the host stars. 
Using these parameters,
we revise orbital and fundamental parameters of 34 planets and four planet candidates.
According to our findings, radius range of the planets is 0.35-16.50 $\rm R_{\earth}$. 
The maximum difference between the transit and revised radii occurs for Kepler-444b-f is about 25 per cent.
\end{abstract}




\section{Introduction}

Planetary studies collect huge data nowadays. Thanks to the {\it Convection, Rotation and planetary Transits} ({\it CoRoT}; \citealt{bag}), {\it Kepler} \citep{koc},
ground-base observations and {\it Transiting Exoplanet Survey Satellite} ({\it TESS}; \citealt{sul}), more than 3900{\footnote{\url{http://exoplanetarchive.ipac.caltech.edu/}}} planets are discovered. 
Fate of a planet is determined and characterized by its host star. 
Accuracy of the fundamental planetary parameters depends on how exact properties of the host stars we compute.
Most of the observed planet and planet candidate host stars have convective envelope. Thus, they exhibit solar-like oscillations. Long-period observations with high precise
data allow to reveal solar-like oscillation frequencies. 
Hereby,
fundamental parameters of host stars are derived from asteroseismic methods 
with the reference frequencies (\citealt{yil14}, hereafter \citetalias{yil14}; \citealt{yil15}, hereafter \citetalias{yil15}) and scaling relations (see e.g. \citealt{mat}; \citealt{hub}). 

From {\it CoRoT} and {\it Kepler} observations, many solar-like oscillating host stars have been discovered. 
First remarkable studies that derive stellar parameters of the host stars in great number using asteroseismology are 
for {\it Kepler} planet-candidate host stars \citep{hub,silva}.
Before {\it Kepler}, some studies that are based on {\it CoRoT}, {\it Hubble} and ground-based observations are seen 
in the literature (see e.g. \citealt{sor}; \citealt{nut}; \citealt{wri}). 
In this study, we construct interior models of 20 {\it Kepler} and {\it CoRoT} target host stars with the {\small MESA} stellar evolution code \citep{pax11,pax13}. 
We compute adiabatic oscillation frequencies of
 the models and compare them with observed oscillation frequencies
with {\small ADIPLS} package
and try to obtain fundamental parameters of the host stars and their planets. 

Since large separation between oscillation frequencies (${\Delta{\nu}}$) is
related to sound travel time throughout stellar radius ($R$), mean stellar density derived from asteroseismology is more accurate 
than the mean density derived from any method \citep{ulr}. 
If effective temperature (${T_{\rm eff}}$) is observed precisely, $R$ and stellar mass ($M$) are determined using the asteroseismic quantities, 
namely, frequency of the maximum amplitude (${\nu_{\rm max}}$), 
{${\Delta{\nu}}$ and reference frequencies (${\nu_{\rm min0}}$, ${\nu_{\rm min1}}$, and ${\nu_{\rm min2}}$; Paper I and II) at the minima of 
${\Delta{\nu}}$}, 
in conventional \citep{kje.bed} and 
 new scaling relations (\citealt{yil16}, hereafter \citetalias{yil16}).
Stellar age is derived from stellar interior models.
As hydrogen fused into helium in the nuclear core, mean molecular weight increases and sound speed gradient changes in time.  
Therefore, the small separation between oscillation frequencies (${\delta{\nu_{02}}}$) is much more sensitive function of the 
nuclear processes in the core than individual oscillation frequencies.
Hence, 
age of main-sequence (MS) stars determined by ${\delta{\nu_{02}}}$ 
is much more precise than ages from any method {\citep{ulr}}.

Most of the planets is discovered by observing their transit across the disk of the hosts  (see e.g. \citealt{ste12}; \citealt{row}). 
Planets 
have also been detected by radial velocity (RV) method.
Although mass of a planet directly estimated from the RV method depends on inclination of its orbit
(see e.g. \citealt{bor10}; \citealt{coc}; \citealt{barc}),
for both methods, properties of the host stars are required for determination of the fundamental planetary parameters.
In this study, we construct interior models for the host stars and obtain their parameters under asteroseismic and non-asteroseismic  constraints.
Then, we revise fundamental and orbital parameters of 34 planets and four planet-candidates 
using model mass and radius of the host stars (see Section~\ref{sec:propplanet}).

We organize this paper as follows. In Section~\ref{sec:prophost}, 
 we give basic properties of the host stars infered from asteroseismic and non-asteroseismic observational data.
{\small MESA} models and seismic analysis of the host stars are presented in Section~\ref{sec:modhost}. 
In Section~\ref{sec:propplanet}, we estimate fundamental and orbital parameters of the planets and planet candidates.
Lastly, we draw our conclusions in Section~\ref{sec:conc}.

\section{ASTEROSEISMIC AND NON-ASTEROSEISMIC PROPERTIES OF THE HOST STARS}
\label{sec:prophost}

The  host stars have different evolutionary phases. They exhibit solar-like oscillations. 
Thus, we have opportunity to analyse the host stars with asteroseismology.  
Observed asteroseismic and spectral properties of the host stars are listed in Table~\ref{tab:observed.prop13host}.
Among the host stars, HD 52265 is the only star observed by {\it CoRoT}. The remaining stars are observed by {\it Kepler}. 
Observational oscillation frequencies of the host stars that are obtained from {\it CoRoT} and {\it Kepler} light curves are taken from the literature
{(see Table 1)}.
From observational oscillation frequencies, we obtain mean small separation between oscillation frequencies (${\braket{\delta{\nu_{02}}}}$).
${\Delta{\nu}}$ and ${\nu_{\rm max}}$ are taken from the literature. 
{${\Delta{\nu}}$ from the literature is in very good agreement with ${\Delta{\nu}}$ obtained from ${\Delta{\nu}}$ versus $\nu$ graph.
Also, we determine the reference frequencies from their ${\Delta{\nu}}$ versus $\nu$ graph. 
For all of the target stars, we have determined ${\nu_{\rm min0}}$ and ${\nu_{\rm min1}}$ from their ${\Delta{\nu}}$ versus $\nu$ graph, except KIC 10963065. 
For KIC 10963065, ${\nu_{\rm min0}}$ is not available.
min2 is partly or entirely seen in the five target stars (KIC 3632418, KIC 8866102, KIC 9592705,
KIC 10666592, and KIC 11807274).

To compute ${\nu_{\rm min}}$ of any minima from ${\Delta{\nu}}$ versus $\nu$ graph,
we first determine
frequency interval of the minimum and draw two straight lines from the neighbourhood
intervals. The intersection of the two lines corresponds ${\nu_{\rm min}}$.

\begin{figure}
\includegraphics[width=\columnwidth]{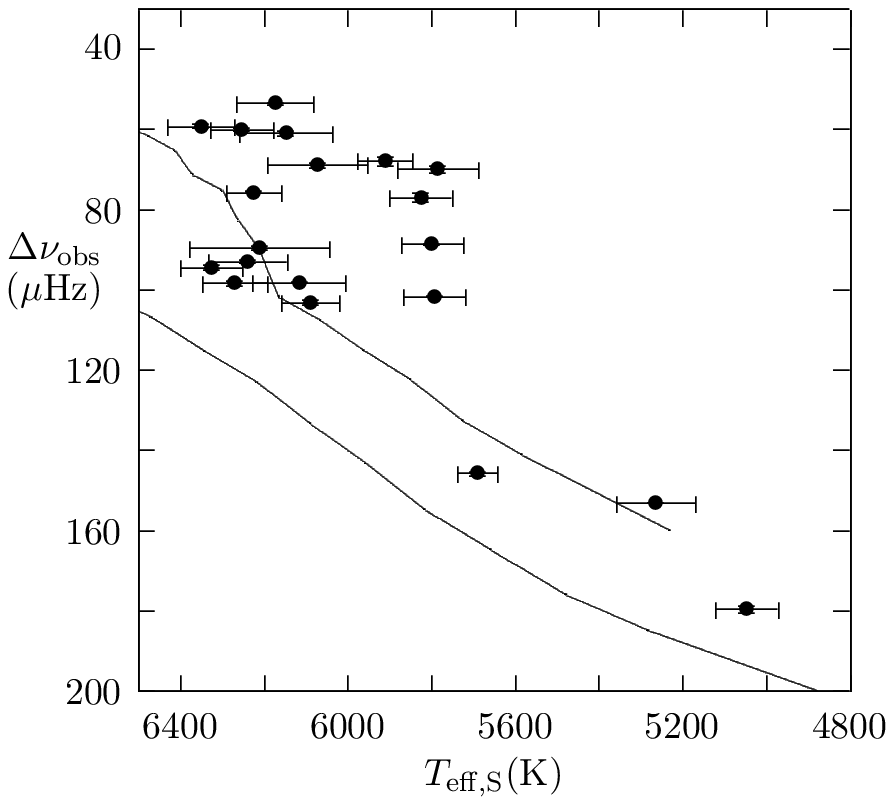}
\caption{Large separation between observed oscillation frequencies with respect to 
{spectroscopic 
effective temperature for the 20 host stars. 
         Thin and thick solid lines are for zero-age main sequence (ZAMS) and terminal-age main sequence (TAMS), respectively.}
}
\label{fig:obs.Dnu.Teff}
\end{figure}

Evolutionary phases of the host stars are seen in Fig.~\ref{fig:obs.Dnu.Teff}. 
In this figure, ${\Delta{\nu}_{\rm obs}}$ of the host stars are plotted with respect to {effective temperature based on spectra (${T_{\rm eff,S}}$)}. 
Thin and thick solid lines are zero-age main sequence (ZAMS) and terminal-age main sequence (TAMS), respectively. 
These lines are computed with the {\small{MESA}} evolution code for the mass range 0.8-1.6 $\rm M_{\sun}$ with solar composition. 
Five host stars are MS star. Most of the host stars are on the subgiant evolutionary phase. 
The effective temperature range of the host stars is about 5000-6350 K. 
KIC~10666592 is the hottest host star and its effective temperature is 6350 $\pm$ 80 K. 
In contrast, effective temperature of the coolest host star (KIC~6278762) is 5046 $\pm$ 74 K.

Most of the 38 planets are small rocky planets. The remaining planets are mostly
hot giant. Number of multiple systems is 10. The planet with the longest period is Kepler-126c, approximately $100$ d. 
Kepler-65c is the planet with the shortest orbital period of~$2$ d. 

\begin{table*}
	\caption{Observed spectral and asteroseismic properties of the host stars. Columns are organized as star name, effective temperature, 
{surface metallicity, spectral and asteroseismic
gravities, frequency of maximum amplitude, mean large and small separations between oscillation frequencies, reference frequencies for min0, min1 and min2 and references.
Effective temperature and observed metallicity are taken from spectral observations. Asteroseismic gravity is computed from asteroseismic scaling relation.
}
${\braket{\delta{\nu_{02}}}}$, ${\nu_{\rm min0}}$, ${\nu_{\rm min1}}$, and ${\nu_{\rm min2}}$ are derived from observed oscillation frequencies. 
${\nu_{\rm max}}$ and ${\Delta{\nu}}$ are taken from the references given in the last column.}
	\label{tab:observed.prop13host}
\begin{center}
\small\addtolength{\tabcolsep}{-3pt}
        \begin{tabular}{lccccccccccr}
                \hline
        Star  & ${T_{\rm eff,S}}$ & ${\rm [M/H]_{obs}}$ &  ${\log g_{\rm s}}$  & ${\log g_{\rm sca }}$&   ${\nu_{\rm max}}$  & $\braket{\Delta{\nu}}$  &  ${\braket{\delta{\nu_{02}}}}$  &
  ${\nu_{\rm min0}}$  &  ${\nu_{\rm min1}}$ &  ${\nu_{\rm min2}}$ & Ref.\\
              &    (K)         &   (dex)         &   (cgs)     &   (cgs)   &      ({$\mu$}Hz)   &      ($\mu$Hz)     &     ($\mu$Hz) &
   ($\mu$Hz)           &     ($\mu$Hz)        &     ($\mu$Hz) & \\
                \hline
        HD~52265     &  6116 $\pm$ 110  &   0.22 $\pm$ 0.05  &  4.32 $\pm$ 0.20 & 4.28 $\pm$ 0.01 & 2090.0 $\pm$  20 &  98.1 $\pm$ 0.1 & 8.2  $\pm$ 0.9 & 2338.1 & 1845.7 &  -- & 3,12,17  \\
        KIC~3544595  &  5689 $\pm$  48  &  -0.15 $\pm$ 0.40  &  4.56 $\pm$ 0.06 & 4.46 $\pm$ 0.02 & 3366.0 $\pm$  81 & 145.8 $\pm$ 0.5 & 8.6  $\pm$ 1.7 & 3283.2 & 2701.9 &  -- & 2,14,23  \\
        KIC~3632418  &  6148 $\pm$ 111  &  -0.19 $\pm$ 0.21  &  3.94 $\pm$ 0.21 & 4.01 $\pm$ 0.04 & 1159.0 $\pm$  44 &  60.9 $\pm$ 0.6 & 4.3  $\pm$ 0.7 & 1422.1 & 1065.2 &   736.0 & 1,9,19,20\\
        KIC~4349452  &  6270 $\pm$  79  &  -0.04 $\pm$ 0.10  &  4.28 $\pm$ 0.03 & 4.29 $\pm$ 0.02 & 2106.0 $\pm$  50 &  98.3 $\pm$ 0.6 & 7.8  $\pm$ 1.6 & 2365.2 & 1884.5 &  -- & 4,14,18  \\
        KIC~5866724  &  6211 $\pm$ 167  &   0.17 $\pm$ 0.06  &  4.23 $\pm$ 0.01 & 4.24 $\pm$ 0.03 & 1880.0 $\pm$  60 &  89.6 $\pm$ 0.5 & 7.6  $\pm$ 1.2 & 2261.4 & 1698.3 &  -- & 8,14	\\
        KIC~6278762  &  5046 $\pm$  74  &  -0.55 $\pm$ 0.07  &  4.60 $\pm$ 0.06 & 4.58 $\pm$ 0.03 & 4538.0 $\pm$ 144 & 179.6 $\pm$ 0.8 & 9.5  $\pm$ 1.0 & 4220.8 & 3411.7 &  -- & 6,10	\\
        KIC~6521045  &  5825 $\pm$  75  &   0.02 $\pm$ 0.10  &  4.13 $\pm$ 0.03 & 4.13 $\pm$ 0.02 & 1502.0 $\pm$  31 &  77.0 $\pm$ 1.1 & 5.3  $\pm$ 0.5 & 1643.2 & 1259.1 &  -- & 10,18	\\
        KIC~7296438  &  5798 $\pm$  75  &   0.30 $\pm$ 0.10  &  4.15 $\pm$ 0.15 & 4.22 $\pm$ 0.01 & 1848.0 $\pm$  16 &  88.7 $\pm$ 0.1 & 5.3  $\pm$ 0.3 & 1983.2 & 1540.8 &  -- & 11,13	\\
        KIC~8077137  &  6072 $\pm$ 121  &  -0.09 $\pm$ 0.15  &  4.07 $\pm$ 0.03 & 4.09 $\pm$ 0.03 & 1324.0 $\pm$  39 &  68.8 $\pm$ 0.6 & 5.6  $\pm$ 1.0 & 1494.3 & 1140.0 &  -- & 10,15	\\
        KIC~8292840  &  6239 $\pm$  94  &  -0.14 $\pm$ 0.10  &  4.25 $\pm$ 0.04 & 4.27 $\pm$ 0.02 & 1983.0 $\pm$  37 &  92.9 $\pm$ 0.4 & 7.8  $\pm$ 1.3 & 2245.6 & 1730.6 &  -- & 10,22	\\
        KIC~8866102  &  6325 $\pm$  75  &   0.01 $\pm$ 0.10  &  --          & 4.28 $\pm$ 0.02 & 2014.0 $\pm$  32 &  94.5 $\pm$ 0.6 & 8.0  $\pm$ 1.7 & 2420.8 & 1801.7 &  1342.0 & 10,24	\\
        KIC~9414417  &  6253 $\pm$  75  &  -0.13 $\pm$ 0.10  &  --          & 4.02 $\pm$ 0.03 & 1115.0 $\pm$  32 &  60.1 $\pm$ 0.3 & 4.5  $\pm$ 1.1 & 1059.6 &  730.2 &  -- & 10	\\
        KIC~9592705  &  6174 $\pm$  92  &   0.22 $\pm$ 0.10  &  --          & 3.97 $\pm$ 0.02 & 1008.0 $\pm$  21 &  53.5 $\pm$ 0.3 & 4.9  $\pm$ 1.2 & 1265.4 &  971.0 &   728.3 & 10	\\
        KIC~9955598  &  5264 $\pm$  95  &   0.08 $\pm$ 0.10  &  4.29 $\pm$ 0.12 & 4.48 $\pm$ 0.03 & 3546.0 $\pm$ 119 & 153.2 $\pm$ 0.1 & 9.0  $\pm$ 0.8 & 3606.2 & 2842.8 &  -- & 1,14,19  \\
        KIC~10514430 &  5784 $\pm$  98  &  -0.11 $\pm$ 0.11  &  --          & 4.07 $\pm$ 0.02 & 1303.0 $\pm$  30 &  70.0 $\pm$ 1.0 & 5.9  $\pm$ 0.7 & 1388.2 & 1006.9 &  -- & 10	\\
        KIC~10666592 &  6350 $\pm$  80  &   0.26 $\pm$ 0.08  &  4.07 $\pm$ 0.08 & 4.02 $\pm$ 0.10 & 1115.0 $\pm$ 110 &  59.2 $\pm$ 0.6 & 4.5  $\pm$ 1.0 & 1569.2 & 1182.2 &   796.5 & 10,21	\\
        KIC~10963065 &  6090 $\pm$  70  &  -0.25 $\pm$ 0.06  &  4.31 $\pm$ 0.08 & 4.29 $\pm$ 0.03 & 2184.0 $\pm$  62 & 103.2 $\pm$ 0.6 & 7.1  $\pm$ 0.9 & 2338.8 & 1817.5 &  -- & 1,5,9,19 \\
        KIC~11295426 &  5793 $\pm$  74  &   0.12 $\pm$ 0.07  &  4.28 $\pm$ 0.06 & 4.27 $\pm$ 0.01 & 2154.0 $\pm$  13 & 101.6 $\pm$ 0.1 & 5.6  $\pm$ 0.8 & 2212.0 & 1767.4 &  -- & 14,16,23 \\
        KIC~11401755 &  5911 $\pm$  66  &  -0.20 $\pm$ 0.06  &  4.05 $\pm$ 0.01 & 4.06 $\pm$ 0.04 & 1250.0 $\pm$  44 &  67.9 $\pm$ 1.2 & 5.2  $\pm$ 1.4 & 1371.0 & 1100.9 &  -- & 7,10	\\
        KIC~11807274 &  6225 $\pm$  66  &   0.06 $\pm$ 0.08  &  4.13 $\pm$ 0.01 & 4.14 $\pm$ 0.04 & 1496.0 $\pm$  56 &  75.7 $\pm$ 0.3 & 8.1  $\pm$ 1.3 & 1680.9 & 1334.7 &   928.6 & 8,14	\\
										  
                \hline
        \end{tabular}

\end{center}
{\it Note.} References: 1: \citet{app}; 2: \citet{bal}; 3: \citet{bal11}; 4: \citet{ben}; 5: \citet{bru}; 6: \citet{cam}; 7: \citet{car}; 8: \citet{cha13}; 9: \citet{cha14};
 10: \citet{dav}; 11: \citet{deh}; 12: \citet{esc}; 13: \citet{eve}; 14: \citet{hub}; 15: \citet{hub14}; 16: \citet{gil13};  17: \citet{lebgou14}; 18: \citet{mar}; 
19: \citet{met14}; 20: \citet{mol}; 21: \citet{pal}; 22: \citet{row}; 23: \citet{san}; and 24: \citet{van}. 
\end{table*}

\section{Interior MODELS OF THE HOST STARS}
\label{sec:modhost}

\subsection{Properties of the {\small{MESA}} code }
We construct interior models of 20 host stars with {\small{MESA}} evolution code \citep{pax11, pax13}.  
Standard mixing length theory \citep{bomv} is used for convection treatment. The effects of convective overshooting are not considered.
OPAL opacity tables are taken from \citet{igl93,igl96}.
In nuclear reaction rates, we use \citet{ang} with updated by \citet{kun} and  \citet{cyb}.
{
Stellar atmospheric conditions are vital for asteroseismic modelling in particular for high-frequency domain. 
For simplicity, we select the {\scriptsize{SIMPLE\_PHOTOSPHERE}} option in {\small{MESA}}  for the host stars (see details in \citealt{pax11}).
}
Element diffusion is included with {\small MESA} default option. 
{Diffusion is taken into account for the host stars with ${M_{\rm star}} < 1.2 \rm~M_{\sun}$.}
For the solar values, initial hydrogen abundance $X=0.70358$, metallicity $Z=0.0172$, age $t=4.57$ {{Gyr}}, and the mixing length parameter ($\alpha=2.175$) are used
for the {\small{MESA}} evolution code. 

We use {\small ADIPLS} package \citep{chris} 
in {\small MESA} module to compute adiabatic oscillation frequencies of interior models. { We compute $\nu_{\rm max}$ from \citet{bro} with 
the solar values}  ($\nu_{\rm max{\sun}}=3050$ $\mu$Hz and $\rm T_{\rm eff{\sun}}=5777$ K). 
{For determination of the reference frequencies}, we apply method in \citetalias{yil14}. Near surface region, because of lower sound speed, 
stellar evolution codes are difficult to simulate. 
Therefore, surface correction is needed. In this study, we apply surface correction in {\small ADIPLS} package (\citealt{kje}). 

\subsection{{Modelling strategy and ${\chi}^2$ method}}
\begin{figure}
\includegraphics[width=\columnwidth]{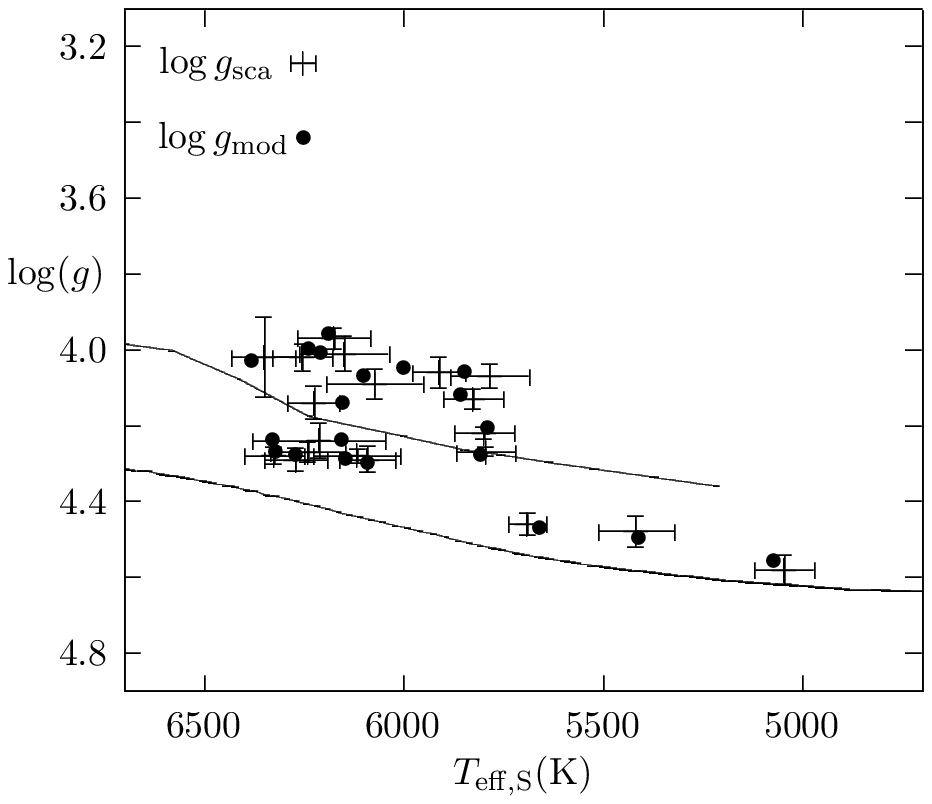}
\caption{
{$\log g_{\rm sca}$  for the 20 host stars is plotted with respect to $T_{\rm eff,S}$ with their uncertainties.
Also plotted is  $\log g_{\rm mod}$ with respect to  $T_{\rm mod}$ of the best-fitting models (filled circles).
         Thin and thick solid lines are for the ZAMS and TAMS lines, respectively, taken from \citet{yi15}.
}
}
\label{fig:logg_sis.krs.Teff}
\end{figure}
{
The input parameters for the {\small MESA} evolution code are $M_{\rm mod}$, initial helium ($Y_{0{\rm mod}}$) and heavy element ($Z_{0{\rm mod}}$) abundances  and $\alpha$. 
Among these parameters, $Z_{0{\rm mod}}$ is derived from the observed metallicity (${\rm [M/H]_{obs}}$) (see below).
For the models with diffusion, however, $Z$ computed from ${\rm [M/H]_{obs}}$ is the metallicity at the surface. $\alpha$ is taken as the solar value.
During the calibration procedure, we 
properly change 
$M_{\rm mod}$ and $Y_{0{\rm mod}}$ in order to fit models to the 
asteroseismic and non-asteroseismic constraints.
If the calibration is not successful, we slightly modify $Z_{0{\rm mod}}$.   
For all of these stars, we have $T_{\rm eff,S}$ from spectra and gravity ($g_{\rm sca}$) from the scaling relation as a function of $T_{\rm eff,S}$ { and $\nu_{\rm max}$:
\begin{equation}
\frac{g_{\rm sca}}{\rm g_{\sun}}=\frac{\nu_{\rm max}}{\nu_{{\rm max}\sun}} \left(\frac{T_{\rm eff,S}}{T_{\sun}}\right)^{0.5}. \nonumber
\end{equation}
}
In Fig.~\ref{fig:logg_sis.krs.Teff}, $\log g_{\rm sca}$ is plotted with respect to $T_{\rm eff,S}$. Also shown are the best-fitting models.  

In fitting interior model of a star to the observational constraints, we first try to fit model of a star to the observed box in the $T_{\rm eff,S}-\log g_{\rm sca}$ diagram and secondly check how asteroseismic
constraints are satisfied by oscillation frequencies of the model. 
The asteroseismic constraints comprise 
the observed oscillation frequencies, 
$\Delta{\nu}$, ${\braket{\delta{\nu_{02}}}}$, ${\nu_{\rm max}}$, and the reference frequencies (${\nu_{\rm min0}}$, ${\nu_{\rm min1}}$, and ${\nu_{\rm min2}}$).
The best-fitting model is decided by applying ${\chi}^2$ method (see below).
We slightly change $\log g_{\rm mod}$ and $T_{\rm mod}$ if needed for the minimization of ${\chi}^2$. 
For all of the host stars, except KIC 9955598, the difference between $T_{\rm eff,S}$ and $T_{\rm mod}$ of the best-fitting model is less than 100 K.

The difference between 
$T_{\rm eff,S}$ and $T_{\rm mod}$ of KIC 9955598 is 148 K.
Effective temperatures of KIC 9955598 computed from its $B-V$ and $V-K$ colours are as 5355 and 5480 K, respectively. Its $T_{\rm mod}$ (5412 K) is very close to the 
mean (5418 K) of the $T_{\rm eff}$s from the colours.
In computation of its ${\chi}^2$, its observed $T_{\rm eff}$ is taken as 5418 K. 

We compute normalized  asteroseismic ${\chi}^2_{\rm seis}$ in order to evaluate resemblance rate between individual oscillation frequencies derived from 
observations and models: 
\begin{equation}
{{\chi^2_{\rm seis}}}={\frac{1}{N_{\rm freq}}}{\sum\limits_{i=1}^{N_{\rm freq}}{\left({\frac{{\nu_{\rm obs}}-{\nu_{\rm mod,i}}}{\sigma_{\rm obs,i}}}\right)^2}},
\label{eq:chisquare.seis}
\end{equation}
where $N_{\rm freq}$, ${\nu_{\rm obs,i}}$, and ${\nu_{\rm mod,i}}$ 
are number of observed oscillation frequencies,  observed and model oscillation frequencies, 
 respectively. ${\sigma_{\rm obs,i}}$ is uncertainty of the observed oscillation frequency.

To fit model oscillation frequencies to observed oscillation frequencies, density is the key parameter (see \citetalias{yil16}).
However, different combinations of $M$ and $R$ may utilize the same mean density but completely different interior models for each of the $M$ and $R$ combinations.
On account of this, we particularly pay attention to use the reference frequencies in our analysis.
The reference frequencies strongly depend on $M$ and $R$. Difference between the best model and observed reference
frequencies for all  of the host stars is in general less than ${\Delta{\nu}}/2$ (see \citetalias{yil14}).}

The observed metallicity is included as input constraints on the models. 
Initial metallicity ($Z_{0{\rm mod}}$) is computed from ${\rm [M/H]_{obs}}$ given in Table 1 using with solar initial metallicity $\rm Z_{0\sun}$:
$Z_{0{\rm mod}}= 10^{\rm [M/H]_{obs}} \rm Z_{0\sun}$. However, the mean uncertainty in ${\rm [M/H]_{obs}}$ of the stars is about 0.11 dex. In addition, 
some extra difficulties in determination of $Z_{0{\rm mod}}$ arise because of differentiation 
of surface metallicity from initial metallicity due to various processes such as diffusion and mixing.
These processes strongly depend on stellar mass and the diffusion is in general not applicable for the stars with  $M>1.20$ M$_{\sun}$.
Therefore, the metallicity is not involved in our non-seismic ${\chi}^2$ analysis.  

In computation of non-seismic ${\chi}^2$, we primarily focus on the observed effective temperature and surface gravity to fit model parameters:
\begin{equation}
{{\chi^2_{\rm spec}}}={{\chi^2_{\rm T_{\rm eff}}}}+{{\chi^2_{\rm \log g}}},
\label{eq:chisquare.spec}
\end{equation}
where
\begin{equation}
{{\chi^2_{\rm T_{\rm eff}}}}={\left({\frac{{T_{\rm eff,S}}-{T_{\rm mod}}}{\sigma_{T_{\rm eff,S}}}}\right)^2}, 
\label{eq:chisquare.teff}
\end{equation}
where $\sigma_{T_{\rm eff,S}}$ is uncertainty in $T_{\rm eff,S}$. 
${T_{\rm mod}}$ is effective temperature of the {\small MESA} model. 

{
For asteroseismic and spectroscopic constraints, ${\chi}^2$ is calculated independently in equation~(\ref{eq:chisquare.seis}) and (\ref{eq:chisquare.spec}),
respectively. 
For ${{\chi^2_{\rm spec}}}$, uncertainties in $T_{\rm eff,S}$ and $\log g$ are involved as in equation~(\ref{eq:chisquare.spec}). 
${{\chi^2_{\log g}}}$ is calculated from an equation similar to equation ~(\ref{eq:chisquare.teff}). 
${{\chi^2_{\rm spec}}}$ of the host stars is listed in Table~\ref{tab:diffevocode.MESA.model}.
}

\begin{figure}
\includegraphics[width=\columnwidth]{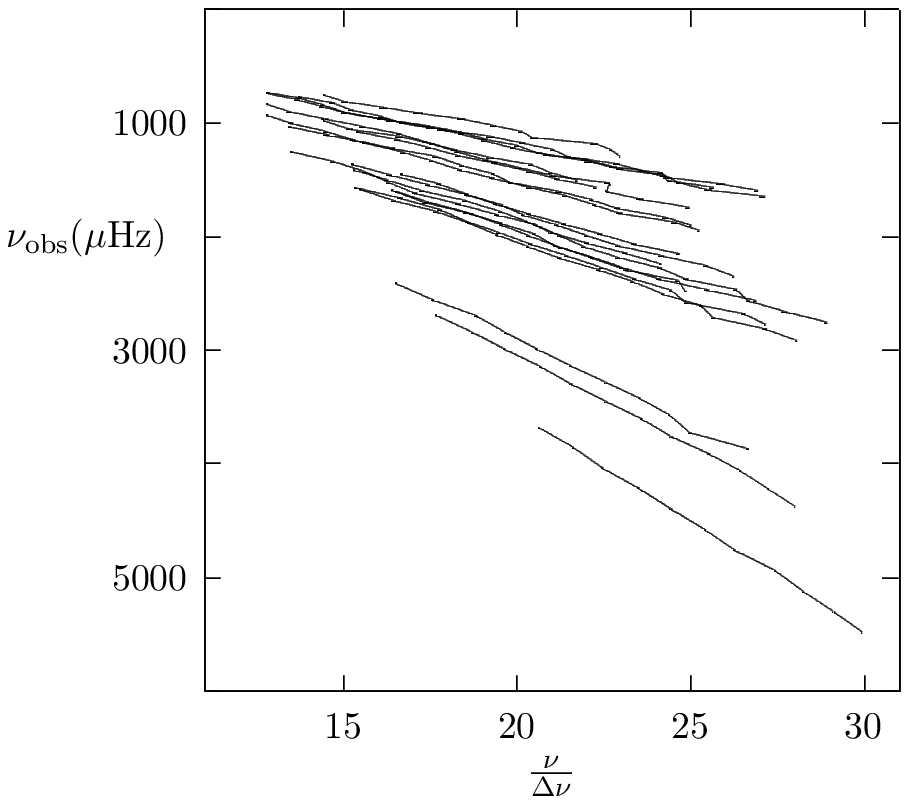}
\caption{Observed oscillation frequencies ($\nu_{\rm obs}$) with $l=0$ are plotted with respect to ${n'_{\rm obs}}$ for 20 planet 
         and planet-candidate 
         host stars. These lines show ${n'_{\rm obs}}$ of each host star. From high to low frequencies, stars evolve from MS to subgiant. 
         KIC~3544595, KIC~9955598, and KIC~6278762 in the high-frequency region are located separately. 
         The remaining stars are on the subgiant phase. They stand in the densest region. They have $700 < \nu_{obs} < 2900$ $\mu$Hz.}
\label{fig:obs.n.nu}
\end{figure}

{
The difference between observed and model oscillation frequencies is in particular significant in the high-frequency domain.
Therefore, we disregard some of the high-frequency data in computation of  ${\chi}^2_{\rm seis}$. 
This difference is due to the observational frequencies.
As long as ${\Delta{\nu}_{\rm obs}}$ is plotted with respect to ${n'_{\rm obs}}=\frac{\nu_{\rm obs}}{\Delta{\nu}_{\rm obs}}$, 
the observed oscillation frequencies of the modes of some dwarfs with ${n'_{\rm obs}} > 25$ are fluctuated. Uncertainties of the frequencies are 
significantly greater than that of the modes with ${n'_{\rm obs}} < 25$ (see also fig. 3 in \citetalias{yil16}).
} 
Actually, this situation depends on evolutionary phase of a star. As it evolves
 from MS to the red giant branch (RGB), range of 
${n'_{\rm obs}}$ is changed. In Fig.~\ref{fig:obs.n.nu}, observed oscillation frequencies with $l=0$ are plotted with respect to 
${n'_{\rm obs}}$ for 20 planet and planet-candidate host stars. 
However, observed oscillation frequencies of low-mass MS star, KIC~6278762, are in $21 < {n'_{\rm obs}} < 30$ range. 
We notice the scattering of data of the modes with the two highest frequencies.
Most of the stars is on the subgiant evolutionary phases, 
oscillation frequencies are observed in 
$13 < {n'_{\rm obs}} < 27$ range. 

Because of nuclear evolution during the MS phase, sound speed gradient changes inside the nuclear core. Change in sound speed gradient causes ${\delta{\nu_{02}}}$ to decrease.
This makes ${\delta{\nu_{02}}}$ a very suitable age indicator for the MS phase. The resolution is very high in a ${\delta{\nu_{02}}}$-${\Delta{\nu}}$ diagram for this phase (see e.g. White et al. 2011).
Beyond the MS phase degeneracy sets in and it seems that  ${\delta{\nu_{02}}}$ provides similar information as ${\Delta{\nu}}$.

As stated above, ${\delta{\nu_{02}}}$ is a very good age indicator for the MS stars. Therefore, observed value of ${\delta{\nu_{02}}}$ is used as one of the 
key constraints for the calibration of models of such stars. For all of the host stars, the difference between observed and model $\braket{\delta{\nu_{02}}}$
is less than 1 $\mu$Hz, except KIC 5866724. For KIC 5866724, the difference is 1.1 $\mu$Hz.

\subsection{Results of the models}

The results of interior model computations for the host stars are listed in Tables 2 and 3.
According to these results, stellar mass range is 0.74-1.55 ${\rm M_{\sun}}$. KIC~10666592 and KIC~6278762 have the highest and the lowest masses, respectively. 
KIC~6278762 also has the lowest radius (0.75 ${\rm R_{\sun}}$) and the oldest stellar age with 11.7 Gyr. 

Most of the host stars have two reference minima
in ${\Delta{\nu}_{\rm obs}}$ versus ${\nu_{\rm obs}}$ graph, 
especially ${\nu_{\rm min0}}$ and ${\nu_{\rm min1}}$. 
Besides these minima, KIC~3632418, KIC~8866102, KIC~9592705, KIC~10666592, and KIC~11807274 entirely or partly have min2.
From the models, ${\nu_{\rm min2}}$ is more stable than ${\nu_{\rm min0}}$ and ${\nu_{\rm min1}}$ for arbitrary mass and abundances. 
High-frequency region in ${\Delta{\nu}}$ versus ${\nu}$ graph is fluctuated. 
Therefore, ${\nu_{\rm min0}}$ is either less or not confidential in some cases. 
Agreement between patterns of observed and model oscillation frequencies in ${\Delta{\nu}}$ versus ${\nu}$ graph reveals the appropriate model parameters.

{In addition to KIC~9955598, we also computed $T_{\rm eff} $ and metallicity of four host stars (KIC~3632418, KIC~10963065, KIC~11295426, and KIC~11807274)
from their colours using \citet{lej} colour and bolometric correction (BC) tables. These results are in agreement with the spectroscopic
results within the uncertainty in general.}
 
{We compare our derived fundamental parameters of the host stars with results obtained by \citet{hub} and \citet{silva}. 
In \citet{hub}, stellar parameters of 66 host stars from asteroseismic constraints are presented.  
\citet{silva} is determined stellar properties of 33 host stars using different grids of stellar evolutionary models.
For ${R_{\rm star}} < 1.3 \rm R_{\sun}$, the agreement between {\small MESA} 
and literature models of the host stars is excellent. However, significant discrepancy between {\small MESA} 
and literature radii  occurs for the range ${R_{\rm star}} > 1.3 \rm R_{\sun}$. 

We also compare our results for stellar mass with the literature. 
Literature masses obtained from \citet{hub} and \citet{silva} ($M_{\rm lit}$) are plotted with respect to stellar mass derived from the {\small MESA} models ($M_{\rm mod}$) 
in Fig.~\ref{fig:replyrefree_M.vs.Mlit}. 
There is mainly a very good agreement 
between $M_{\rm lit}$ and $M_{\rm mod}$, especially
for the masses lower than 1.1 $\rm M_{\sun}$. The significant difference appears for few stars. 

\begin{figure}
\includegraphics[width=\columnwidth]{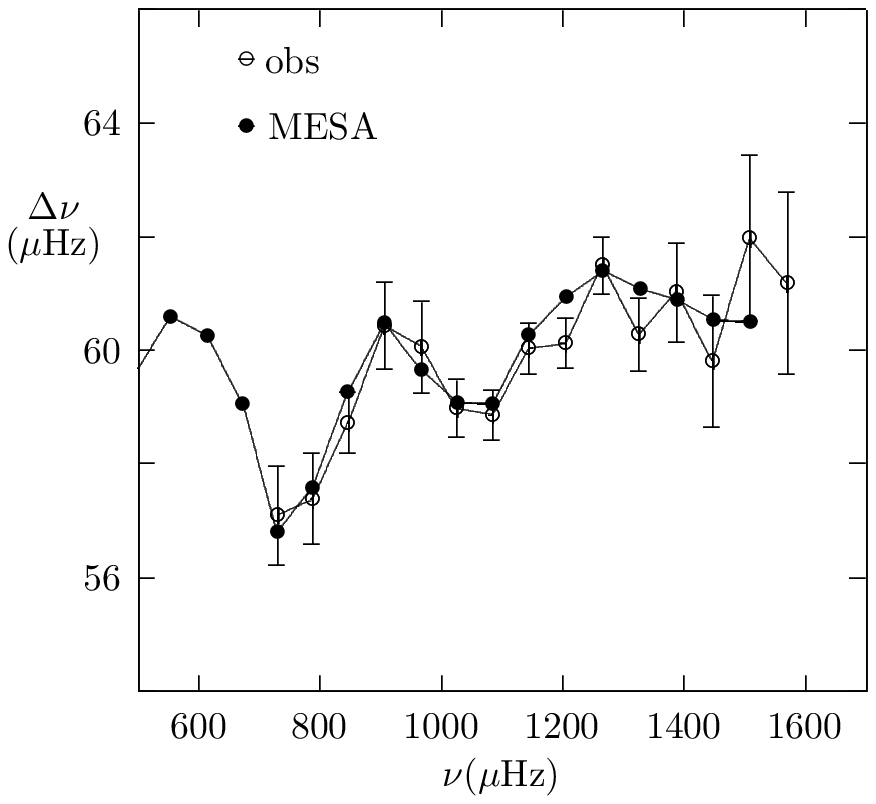}
\caption{Comparison of {\small MESA} model (filled circle) and observed (circle) oscillation frequencies with $l=0$ of KIC 9414417
in ${\Delta{\nu}}$ versus ${\nu}$ graph.}
\label{fig:Dnu.nu.KIC9414417}
\end{figure}

\begin{figure}
\includegraphics[width=\columnwidth]{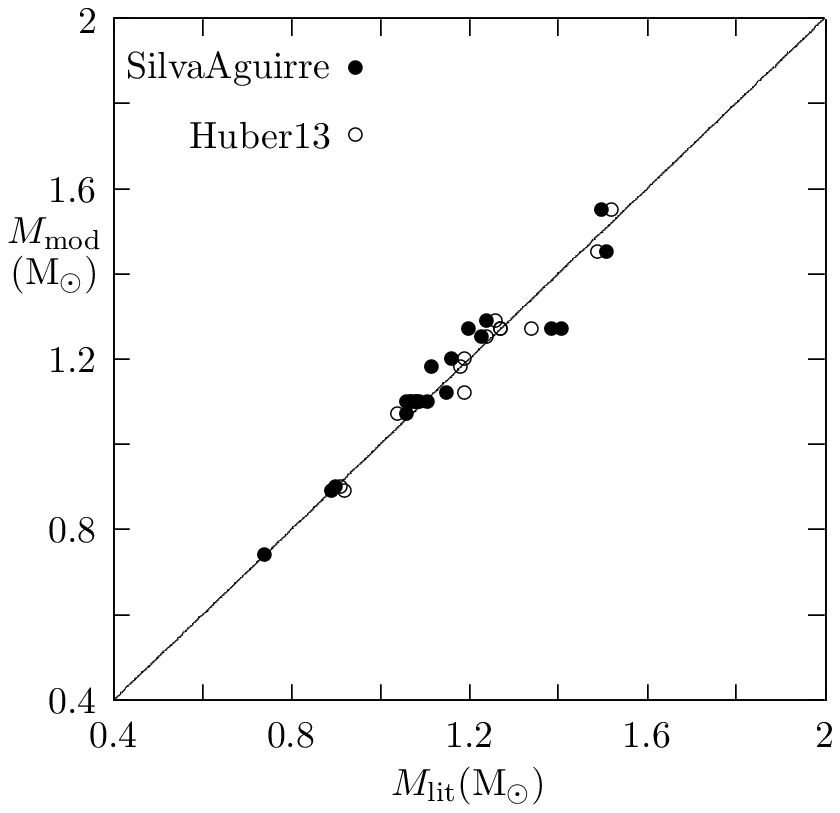}
\caption{{Stellar mass obtained by \citet{hub} and \citet{silva} ($M_{\rm lit}$) is plotted with respect to stellar mass derived from models 
($M_{\rm mod}$) in solar unit. 
}}
\label{fig:replyrefree_M.vs.Mlit}
\end{figure}

\begin{table*}
        \caption{Fundamental {\small MESA} model parameters of the host stars. $M_{\rm mod}$, $R_{\rm mod}$, ${T_{\rm mod}}$, $L_{\rm mod}$, ${\log g_{\rm mod}}$, $Y_{\rm 0{\rm mod}}$, $Z_{0{\rm mod}}$, 
                 and $t_{\rm mod}$ are, 
                 respectively, stellar mass in $\rm M_{\sun}$ unit, stellar radius in $\rm R_{\sun}$ unit, effective temperature in K unit, luminosity in $\rm L_{\sun}$ unit, logarithm of surface 
                 gravity of the model in cgs, and age in units of Gyr. ${{\chi^2_{\rm spec}}}$ of the models  is in the last column.}
        \label{tab:diffevocode.MESA.model}
        \begin{tabular}{lccccccccr}
                \hline
         Star   & $M_{\rm mod}$  & $R_{\rm mod}$ & ${T_{\rm mod}}$ & $L_{\rm mod}$ & ${\log g_{\rm mod}}$ & $Y_{0{\rm mod}}$ & $Z_{0{\rm mod}}$ &
          $t_{\rm mod}$ & ${{\chi^2_{\rm spec}}}$ \\  
              & $(\rm M_{\sun})$ & $(\rm R_{\sun})$ & (K) & $(\rm L_{\sun})$ &  (cgs) &  &  & (Gyr) &  \\ 
                \hline
HD      52265&  1.23 $\pm$  0.04&  1.32 $\pm$  0.01&  6144 $\pm$   110&  2.24 $\pm$  0.16&  4.29 $\pm$  0.04& 0.274 $\pm$ 0.027& 0.023 $\pm$ 0.002&   3.1 $\pm$   0.5&   0.1  \\
KIC   3544595&  0.90 $\pm$  0.10&  0.92 $\pm$  0.04&  5658 $\pm$    48&  0.77 $\pm$  0.07&  4.47 $\pm$  0.14& 0.292 $\pm$ 0.097& 0.015 $\pm$ 0.005&   6.3 $\pm$   2.2&   2.7  \\
KIC   3632418&  1.27 $\pm$  0.09&  1.85 $\pm$  0.05&  6208 $\pm$   111&  4.55 $\pm$  0.41&  4.01 $\pm$  0.09& 0.283 $\pm$ 0.060& 0.018 $\pm$ 0.004&   3.9 $\pm$   1.6&   0.4  \\
KIC   4349452&  1.20 $\pm$  0.05&  1.31 $\pm$  0.02&  6270 $\pm$    79&  2.39 $\pm$  0.14&  4.28 $\pm$  0.05& 0.279 $\pm$ 0.035& 0.017 $\pm$ 0.002&   2.7 $\pm$   0.9&   0.0  \\
KIC   5866724&  1.27 $\pm$  0.06&  1.42 $\pm$  0.02&  6155 $\pm$   167&  2.61 $\pm$  0.29&  4.24 $\pm$  0.06& 0.274 $\pm$ 0.039& 0.022 $\pm$ 0.003&   2.9 $\pm$   0.8&   1.1  \\
KIC   6278762&  0.74 $\pm$  0.03&  0.75 $\pm$  0.01&  5072 $\pm$    74&  0.33 $\pm$  0.02&  4.56 $\pm$  0.05& 0.284 $\pm$ 0.035& 0.012 $\pm$ 0.001&  11.7 $\pm$   2.7&   0.6  \\
KIC   6521045&  1.10 $\pm$  0.05&  1.51 $\pm$  0.03&  5856 $\pm$    75&  2.41 $\pm$  0.16&  4.12 $\pm$  0.06& 0.277 $\pm$ 0.038& 0.019 $\pm$ 0.003&   7.3 $\pm$   1.0&   0.3  \\
KIC   7296438&  1.13 $\pm$  0.04&  1.39 $\pm$  0.02&  5790 $\pm$    75&  1.96 $\pm$  0.12&  4.20 $\pm$  0.05& 0.261 $\pm$ 0.028& 0.023 $\pm$ 0.002&   7.6 $\pm$   0.7&   0.5  \\
KIC   8077137&  1.18 $\pm$  0.07&  1.66 $\pm$  0.04&  6099 $\pm$   121&  3.43 $\pm$  0.32&  4.07 $\pm$  0.08& 0.279 $\pm$ 0.050& 0.017 $\pm$ 0.003&   5.2 $\pm$   1.3&   0.1  \\
KIC   8292840&  1.12 $\pm$  0.05&  1.33 $\pm$  0.02&  6328 $\pm$    94&  2.54 $\pm$  0.17&  4.24 $\pm$  0.05& 0.289 $\pm$ 0.039& 0.012 $\pm$ 0.002&   3.5 $\pm$   0.7&   1.0  \\
KIC   8866102&  1.25 $\pm$  0.05&  1.36 $\pm$  0.02&  6320 $\pm$    75&  2.66 $\pm$  0.15&  4.27 $\pm$  0.05& 0.278 $\pm$ 0.033& 0.019 $\pm$ 0.002&   2.3 $\pm$   0.7&   0.0  \\
KIC   9414417&  1.27 $\pm$  0.05&  1.86 $\pm$  0.03&  6236 $\pm$    75&  4.72 $\pm$  0.27&  4.00 $\pm$  0.05& 0.280 $\pm$ 0.033& 0.016 $\pm$ 0.002&   3.9 $\pm$   1.7&   0.1  \\
KIC   9592705&  1.45 $\pm$  0.06&  2.10 $\pm$  0.03&  6187 $\pm$    92&  5.80 $\pm$  0.38&  3.96 $\pm$  0.05& 0.270 $\pm$ 0.034& 0.026 $\pm$ 0.003&   3.3 $\pm$   0.9&   0.0  \\
KIC   9955598&  0.89 $\pm$  0.04&  0.88 $\pm$  0.01&  5412 $\pm$    95&  0.60 $\pm$  0.04&  4.50 $\pm$  0.05& 0.280 $\pm$ 0.038& 0.017 $\pm$ 0.002&   8.2 $\pm$   2.2&   3.1  \\
KIC  10514430&  1.07 $\pm$  0.05&  1.59 $\pm$  0.03&  5846 $\pm$    98&  2.66 $\pm$  0.20&  4.06 $\pm$  0.06& 0.280 $\pm$ 0.039& 0.016 $\pm$ 0.002&   7.5 $\pm$   1.7&   0.4  \\
KIC  10666592&  1.55 $\pm$  0.15&  1.99 $\pm$  0.07&  6381 $\pm$    80&  5.92 $\pm$  0.51&  4.03 $\pm$  0.12& 0.281 $\pm$ 0.082& 0.025 $\pm$ 0.007&   1.9 $\pm$   0.5&   0.4  \\
KIC  10963065&  1.10 $\pm$  0.04&  1.23 $\pm$  0.02&  6090 $\pm$    70&  1.88 $\pm$  0.11&  4.30 $\pm$  0.05& 0.272 $\pm$ 0.030& 0.014 $\pm$ 0.002&   4.0 $\pm$   1.2&   0.0  \\
KIC  11295426&  1.10 $\pm$  0.03&  1.25 $\pm$  0.01&  5807 $\pm$    74&  1.60 $\pm$  0.09&  4.28 $\pm$  0.03& 0.273 $\pm$ 0.022& 0.024 $\pm$ 0.002&   6.8 $\pm$   1.3&   0.0  \\
KIC  11401755&  1.10 $\pm$  0.05&  1.64 $\pm$  0.04&  5998 $\pm$    66&  3.12 $\pm$  0.20&  4.05 $\pm$  0.07& 0.283 $\pm$ 0.039& 0.013 $\pm$ 0.002&   5.9 $\pm$   3.1&   1.7  \\
KIC  11807274&  1.29 $\pm$  0.06&  1.60 $\pm$  0.02&  6154 $\pm$    66&  3.31 $\pm$  0.16&  4.14 $\pm$  0.05& 0.277 $\pm$ 0.039& 0.020 $\pm$ 0.003&   3.5 $\pm$   0.7&   2.2  \\
               \hline
        \end{tabular}
\end{table*}

\begin{table*}
        \caption{{Asteroseismic parameters of {\small MESA} models for the host stars}. 
                 ${\braket{{\delta{\nu_{02,\rm mod}}}}}$, ${\braket{\Delta{\nu_{\rm mod}}}}$, ${\nu_{\rm max,mod}}$, 
                 ${\nu_{\rm min0,mod}}$, ${\nu_{\rm min1,mod}}$, and ${\nu_{\rm min2,mod}}$ are, respectively, mean small and large separation between model oscillation frequencies,
                 model frequency of maximum amplitude, reference frequencies of model in $\mu$Hz units. ${\nu_{\rm max,mod}}$ is computed from scaling relations with 
                 ${T_{\rm mod}}$ and ${\log g_{\rm mod}}$. 
                 {Typical uncertainties for the reference frequencies are ${\braket{\Delta{\nu_{\rm mod}}}}/2$.}}
        \label{tab:diff.MESA}
        \begin{tabular}{lcrrrrr}
                \hline
         Star   &  ${\braket{{\delta{\nu_{02,\rm mod}}}}}$ & ${\braket{\Delta{\nu_{\rm mod}}}}$ & ${\nu_{\rm max,mod}}$ & ${\nu_{\rm min0,mod}}$ &  
          ${\nu_{\rm min1,mod}}$ & ${\nu_{\rm min2,mod}}$ \\
         &  ($\mu$Hz) & ($\mu$Hz) & ($\mu$Hz) & ($\mu$Hz) & ($\mu$Hz) & ($\mu$Hz) \\
                \hline
        HD~52265   & 7.5 &  98.8 & 2087.8 & 2398.0 & 1857.1 & 1340.9 \\
       KIC~3544595 & 8.8 & 146.5 & 3305.8 & 3286.7 & 2702.4 & 2034.0 \\
       KIC~3632418 & 4.5 &  60.9 & 1096.5 & 1473.3 & 1065.5 &  762.4 \\
       KIC~4349452 & 7.9 &  98.4 & 2040.9 & 2488.8 & 1884.5 & 1397.0 \\
       KIC~5866724 & 6.5 &  89.8 & 1855.8 & 2174.3 & 1633.6 & 1223.3 \\
       KIC~6278762 & 8.5 & 180.6 & 4305.2 & 4221.5 & 3324.0 & 2606.5 \\
       KIC~6521045 & 5.0 &  77.4 & 1461.5 & 1647.1 & 1259.7 &  894.3 \\
       KIC~7296438 & 5.2 &  89.1 & 1781.9 & 1950.9 & 1489.5 & 1094.7 \\
       KIC~8077137 & 4.9 &  69.4 & 1271.1 & 1594.6 & 1148.3 &  847.3 \\
       KIC~8292840 & 6.9 &  93.1 & 1856.3 & 2344.6 & 1742.8 & 1274.7 \\
       KIC~8866102 & 7.6 &  94.5 & 1964.9 & 2521.7 & 1830.8 & 1342.0 \\
       KIC~9414417 & 4.6 &  59.8 & 1074.2 & 1066.3 &  742.5 &  433.5 \\
       KIC~9592705 & 4.2 &  54.0 &  970.0 & 1298.6 &  963.7 &  670.4 \\
       KIC~9955598 & 8.1 & 154.1 & 3621.6 & 3609.3 & 2842.4 & 2228.4 \\
       KIC~10514430 & 5.2 &  70.4 & 1278.4 & 1385.5 & 1077.4 &  800.4 \\
       KIC~10666592 & 4.4 &  59.2 & 1132.5 & 1602.6 & 1183.6 &  829.3 \\
       KIC~10963065 & 7.4 & 103.4 & 2145.9 & 2394.4 & 1859.2 & 1376.9 \\
       KIC~11295426 & 5.2 & 101.8 & 2130.6 & 2238.6 & 1769.2 & 1364.8 \\
       KIC~11401755 & 5.0 &  68.0 & 1228.7 & 1371.9 & 1081.6 &  796.8 \\
       KIC~11807274 & 5.5 &  75.9 & 1483.5 & 1836.4 & 1336.7 & 1002.5 \\
                \hline
        \end{tabular}
\end{table*}

\subsection{Uncertainties in model parameters}
The uncertainties in $M_{\rm mod}$, $R_{\rm mod}$, and age given in Table 2 are computed using the method obtained by Bellinger (2019). 
This method is a comprehensive method and developed for the MS and early subgiant stars.
In this method, uncertainties of $M_{\rm mod}$ and $R_{\rm mod}$ are computed from uncertainties of 
${T_{\rm mod}}$, metallicity, $\braket{\Delta{\nu}}$, and ${\nu_{\rm max}}$.
For the uncertainty in age, in addition to these parameters, ${\braket{\delta{\nu_{02}}}}$ is also included.
Uncertainties of $L_{\rm mod}$ and ${\log g_{\rm mod}}$ are derived from the uncertainties of $M_{\rm mod}$, $R_{\rm mod}$, and $T_{\rm mod}$
with a quadratic approach.
For $L_{\rm mod}$, for example, 
\begin{equation}
\frac{\Delta L_{\rm mod}}{L_{\rm mod}}=\sqrt{ \left(2\frac{\Delta R_{\rm mod}}{R_{\rm mod}}\right)^2+
\left(4\frac{\Delta T_{\rm mod}}{T_{\rm mod}}\right)^2}.
\end{equation}
In the computations, $\Delta T_{\rm mod}$ is taken the same as $\Delta T_{\rm eff,S}$.
All stars of the 20 hosts are either MS or early subgiant stars.

The uncertainty in $Y_{\rm 0{\rm mod}}$ mostly depends on ${\Delta M_{\rm mod}}$ because we consider 
$Y_{\rm 0{\rm mod}}$ and ${\Delta M_{\rm mod}}$ as variable in order to fit model to the observed luminosity. Using the numerical logarithmic derivatives of model luminosity with respect 
to ${\Delta M_{\rm mod}}$ and $Y_{\rm 0{\rm mod}}$, we obtain
$\Delta Y_{\rm mod}/Y_{\rm mod}\approx 3{\Delta M_{\rm mod}}/{M_{\rm mod}}$.
We apply a similar method for uncertainty in $Z_{\rm mod}$.

\section{Mass and Radius Estimation of the Planets and Planet Candidates}
\label{sec:propplanet}

Most of the planets are discovered by transit method. Besides the method, RV is an important tool for non-transiting planetary systems. 
Confirmed planets by these methods highly depend on fundamental parameters of host stars. Especially, accuracy of stellar radius and mass is crucial. 
We obtain fundamental parameters of the hosts by constructing interior models. In this section,
we compute basic properties of 34 planets and also four planet candidates using these stellar parameters. Then, fundamental orbital and structure parameters of the transiting planets and 
planet candidates are revised. 

30 planets of the host stars are detected by transit method. We revise radius, semimajor axis, and inclination of these planets in this study. 
Radii of the planets are computed using estimated stellar radius ($R_{\rm mod}$) and observed transit data from \citet{row15}. 
Semimajor axis of the planets is estimated using $M_{\rm mod}$.
We also derive orbital inclination of the transiting planets. The orbital inclination is computed from equation (13) in \citet{sea}. 
In that equation, we use $R_{\rm mod}$ and semimajor axis 
and also impact parameter. The impact parameter is taken from \citet{row15}. 

Estimated transiting planetary radius, semimajor axis, and orbital inclination are listed in Table~\ref{tab:planeTR.MESA}. Radius range of the planets is 0.35-16.50 $\rm R_{\earth}$. 
Orbital inclination of the planets is approximately 90{\degr}. Semimajor axis range of the planets is
0.04-0.35 au. According to values of the semimajor axis and radius, Kepler-2b is classified as hot Jupiter. 
We also derive fundamental parameters of the planet candidates.  
These parameters are also derived for transiting planet candidates 
and listed in Table~\ref{tab:planecandidateTR.MESA}. The planet-candidates radius range is 0.55-3.15 $\rm R_{\earth}$. 

{In Fig.~\ref{fig:Rplitgez}, fractional difference between the transit and revised radii of the planets and planet candidates (${{\Delta{R_{\rm p}}}/R_{\rm p}}$) 
is plotted with respect to orbital period (${P_{\rm orb}}$) in units of day. 
This figure shows that the maximum difference between estimated and transit radii is  about 25 per cent for Kepler-444 system.

We estimate the planetary mass for only the planets detected by RV method. 
Among the planets we consider, only four of them (HD~52265b, Kepler-25d, Kepler-68d, and Kepler-93c)
have RV data.
Estimated stellar masses are used in equation (1) in \citet{lebgou14} to obtain the planetary mass. 
Orbital period, eccentricity, inclination of planetary orbit, and semi-amplitude are taken from the literature. 
If eccentricity is not available in the literature, we assume that the orbit is circular. 
For the systems with unknown orbital inclination, we present the minimum mass ($M \sin i$).
The estimated mass is plotted with respect to planetary mass from RV data in Fig.~\ref{fig:Mplitrv}.
Mass range of the planets is 0.95-3 $\rm M_{\rm jup}$.
Kepler-25d and Kepler-93c have the lowest and the highest masses, respectively.         

The estimated planetary mass of HD~52265b is 16 per cent greater than the 
mass from the literature.
Masses of Kepler-25d and Kepler-93c are estimated as 0.29 and 2.98 $\rm M_{\rm jup}$, respectively.   
Updated mass of the non-transiting planets are listed in Table~\ref{tab:planeRV.MESA}.

\begin{figure}
\includegraphics[width=\columnwidth]{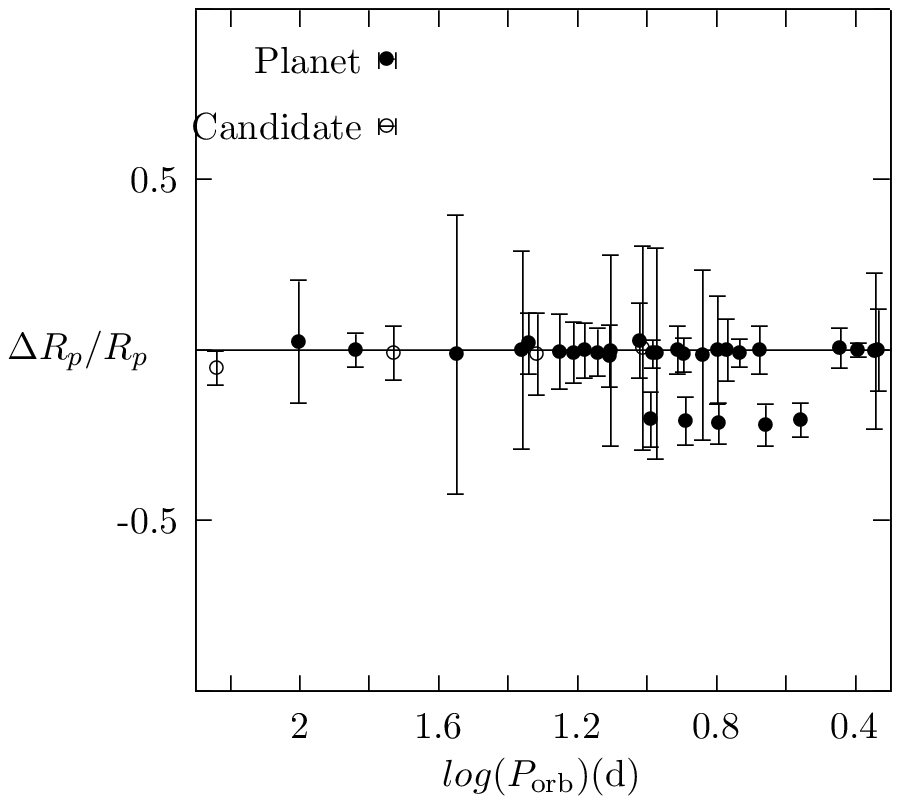}
\caption{{ Fractional radius difference between transit and revised radius of the planets (filled circle) 
and planet candidates (circle) is plotted with respect to orbital period (${P_{\rm orb}}$) in days unit. 
{${{\Delta{R_{\rm p}}}/R_{\rm p}}$} is equal to {$({R_{\rm p,transit}}-{R_{\rm p,mod}})/{R_{\rm p,mod}}$}. 
Transit radius of the planet and planet candidates is taken from \citet{row15}.}} 
\label{fig:Rplitgez}
\end{figure}

\begin{table*}
        \caption{Properties of the transiting planets. Planetary name, orbital period ($P$), radius 
                 [from the literature ($R_{\rm plit}$) and this study ($R_{\rm p}$)], semimajor axis ($a$), and inclination ($i$) of the planetary orbit are presented. 
                 $a$ and $i$ are derived from this study. $P$ and $R_{\rm plit}$ are taken from \citet{row15}.}
        \label{tab:planeTR.MESA}
        \begin{tabular}{lrrrcc}
                \hline
       Planet   & $P$ $ $ $ $ $ $  $ $ $ $ $ $ $ $ $ $  $ $ $ $ $ $ $ $ $ $  & $R_{\rm plit}$ $ $ $ $ $ $ $ $ $ $ & $R_{\rm p}$ $ $ & $a$ & $i$  \\
                & (d) $ $ $ $ $ $ $ $  $ $ $ $ $ $ $ $  $ $ $ $ $ $ $ $   & ($\rm R_{\earth}$) $ $ $ $ $ $ $ $   & ($\rm R_{\earth}$) & (au)  & (\degr) \\ 
                \hline
Kepler-2b    &   2.204735 $\pm$  0.000000 & 16.39 $\pm$  0.15 &   16.42 $\pm$ 0.08 &   0.04 $\pm$ 0.01 &  90.00 \\
Kepler-21b   &   2.785822 $\pm$  0.000004 &  1.59 $\pm$  0.01 &    1.58 $\pm$ 0.05 &   0.04 $\pm$ 0.01 &  84.95 \\
Kepler-25b   &  12.720374 $\pm$  0.000002 &  4.51 $\pm$  0.08 &    4.52 $\pm$ 0.20 &   0.11 $\pm$ 0.01 &  89.66 \\
Kepler-25c   &   6.238535 $\pm$  0.000002 &  2.64 $\pm$  0.04 &    2.64 $\pm$ 0.12 &   0.07 $\pm$ 0.01 &  89.95 \\
Kepler-36b   &  16.231920 $\pm$  0.000014 &  3.94 $\pm$  0.05 &    3.97 $\pm$ 0.04 &   0.13 $\pm$ 0.01 &  87.61 \\
Kepler-36c   &  13.849843 $\pm$  0.000059 &  1.48 $\pm$  0.02 &    1.49 $\pm$ 0.05 &   0.12 $\pm$ 0.01 &  89.18 \\
Kepler-50b   &   7.812858 $\pm$  0.000020 &  1.54 $\pm$  0.03 &    1.56 $\pm$ 0.02 &   0.08 $\pm$ 0.01 &  88.27 \\
Kepler-50c   &   9.376643 $\pm$  0.000019 &  1.82 $\pm$  0.03 &    1.84 $\pm$ 0.28 &   0.10 $\pm$ 0.01 &  87.34 \\
Kepler-65b   &   5.859939 $\pm$  0.000003 &  2.55 $\pm$  0.04 &    2.55 $\pm$ 0.05 &   0.07 $\pm$ 0.01 &  89.84 \\
Kepler-65c   &   2.154909 $\pm$  0.000002 &  1.50 $\pm$  0.03 &    1.50 $\pm$ 0.09 &   0.04 $\pm$ 0.01 &  83.02 \\
Kepler-65d   &   8.131225 $\pm$  0.000014 &  1.76 $\pm$  0.03 &    1.76 $\pm$ 0.04 &   0.09 $\pm$ 0.01 &  86.15 \\
Kepler-68b   &   5.398754 $\pm$  0.000002 &  2.29 $\pm$  0.03 &    2.31 $\pm$ 0.01 &   0.06 $\pm$ 0.01 &  89.89 \\
Kepler-68c   &   9.605039 $\pm$  0.000032 &  0.92 $\pm$  0.01 &    0.93 $\pm$ 0.03 &   0.09 $\pm$ 0.01 &  89.74 \\
Kepler-93b   &   4.726740 $\pm$  0.000002 &  1.59 $\pm$  0.04 &    1.59 $\pm$ 0.03 &   0.05 $\pm$ 0.01 &  86.88 \\
Kepler-100b  &  12.815884 $\pm$  0.000018 &  2.28 $\pm$  0.06 &    2.32 $\pm$ 0.03 &   0.11 $\pm$ 0.01 &  87.85 \\
Kepler-100c  &   6.887060 $\pm$  0.000020 &  1.31 $\pm$  0.03 &    1.33 $\pm$ 0.22 &   0.07 $\pm$ 0.01 &  87.19 \\
Kepler-100d  &  35.333087 $\pm$  0.000216 &  1.50 $\pm$  0.03 &    1.52 $\pm$ 0.38 &   0.22 $\pm$ 0.01 &  88.91 \\
Kepler-126b  &  10.495678 $\pm$  0.000017 &  1.54 $\pm$  0.02 &    1.50 $\pm$ 0.09 &   0.10 $\pm$ 0.01 &  87.50 \\
Kepler-126c  & 100.282869 $\pm$  0.000174 &  2.47 $\pm$  0.04 &    2.41 $\pm$ 0.14 &   0.44 $\pm$ 0.06 &  89.98 \\
Kepler-126d  &  21.869676 $\pm$  0.000054 &  1.56 $\pm$  0.03 &    1.53 $\pm$ 0.06 &   0.16 $\pm$ 0.02 &  88.40 \\
Kepler-128b  &  15.089602 $\pm$  0.000044 &  1.43 $\pm$  0.03 &    1.43 $\pm$ 0.05 &   0.13 $\pm$ 0.01 &  89.93 \\
Kepler-128c  &  22.802981 $\pm$  0.000108 &  1.42 $\pm$  0.03 &    1.42 $\pm$ 0.26 &   0.17 $\pm$ 0.01 &  88.30 \\
Kepler-408b  &   2.465026 $\pm$  0.000005 &  0.70 $\pm$  0.01 &    0.70 $\pm$ 0.01 &   0.04 $\pm$ 0.01 &  85.89 \\
Kepler-409b  &  68.958608 $\pm$  0.000214 &  0.98 $\pm$  0.02 &    0.98 $\pm$ 0.03 &   0.32 $\pm$ 0.01 &  89.90 \\
Kepler-410Ab &  17.833682 $\pm$  0.000012 &  2.47 $\pm$  0.04 &    2.48 $\pm$ 0.07 &   0.14 $\pm$ 0.01 &  90.00 \\
Kepler-444b  &   3.600106 $\pm$  0.000008 &  0.31 $\pm$  0.03 &    0.39 $\pm$ 0.02 &   0.04 $\pm$ 0.01 &  88.27 \\
Kepler-444c  &   4.545878 $\pm$  0.000007 &  0.39 $\pm$  0.04 &    0.50 $\pm$ 0.02 &   0.05 $\pm$ 0.01 &  88.60 \\
Kepler-444d  &   6.189406 $\pm$  0.000013 &  0.40 $\pm$  0.04 &    0.51 $\pm$ 0.02 &   0.06 $\pm$ 0.01 &  89.13 \\
Kepler-444e  &   7.743476 $\pm$  0.000017 &  0.42 $\pm$  0.04 &    0.53 $\pm$ 0.03 &   0.07 $\pm$ 0.01 &  88.99 \\
Kepler-444f  &   9.740484 $\pm$  0.000014 &  0.51 $\pm$  0.05 &    0.64 $\pm$ 0.03 &   0.08 $\pm$ 0.01 &  89.75 \\
                \hline
        \end{tabular}
\end{table*}

\begin{table*}
        \caption{Properties of the transiting planet candidates. Columns are organized same as in Table~\ref{tab:planeTR.MESA}.}
        \label{tab:planecandidateTR.MESA}
        \begin{tabular}{lrclccl}
                \hline
     $ $ $ $ $ $  Planet candidate  &  $P$   $ $   $ $ $ $ $ $  $ $ $ $ $ $  $ $ $ $ $ $ $ $ $ $ $ $ & $R_{\rm plit}$ & $ $ $ $ $R_{\rm p}$ & $a$ & $i$ \\
         & (d)       $ $  $ $ $ $ $ $  $ $ $ $ $ $ $ $ $ $ $ $ & ($\rm R_{\earth}$)  $ $ $ $ & ($\rm R_{\earth}$) & (au) & (\degr) \\ 
                \hline
KOI-263      &  20.719520 $\pm$  0.000062 &  2.32 $\pm$  0.05 &    2.35 $\pm$ 0.07 &   0.15 $\pm$ 0.01 &  89.92 \\
KOI-288      &  10.275317 $\pm$  0.000012 &  3.17 $\pm$  0.06 &    3.15 $\pm$ 0.24 &   0.11 $\pm$ 0.01 &  87.06 \\
KOI-364      & 173.877461 $\pm$  0.000000 &  0.54 $\pm$  0.03 &    0.57 $\pm$ 0.02 &   0.63 $\pm$ 0.03 &  89.93 \\
KOI-974      &  53.505838 $\pm$  0.000149 &  2.49 $\pm$  0.06 &    2.51 $\pm$ 0.02 &   0.30 $\pm$ 0.02 &  89.98 \\
                \hline
        \end{tabular}
\end{table*}

\begin{table*}
        \caption{Properties of the non-transiting planets. Planetary name, orbital properties of the planets [period ($P$), semimajor axis ($a$), and 
                 eccentricity ($e$)], planetary mass [from the literature ($M_{\rm plit}$) and this study ($M_{\rm p}$)] and references are presented. 
                 $M_{\rm spe}$ denotes the species of mass. $P$, $a$, $e$, and $M_{\rm spe}$ are taken from the literature.}
        \label{tab:planeRV.MESA}
        \begin{tabular}{lcccclcl}
                \hline
         Planet & $P$ & $a$ & $e$ & $M_{\rm plit}$ & $M_{\rm spe}$ & $M_{\rm p}$ & Ref. \\
                &  (d)   & (au)   & & ($\rm M_{\rm jup}$) & & ($\rm M_{\rm jup}$) & \\ 
                \hline
HD~52265b   &    119.60 $\pm$  0.42 & 0.50 $\pm$ 0.35 &  0.03            & 1.05 $\pm$  0.03 &  $M\sin i$ &  1.08 $\pm$ 0.03 & \citet{nae}   \\
Kepler-25d  &    123.00 $\pm$  2.00 &  --            &        --       & 0.28 $\pm$  0.04 &  $M$     &  0.29 $\pm$ 0.05 & \citet{mar}   \\
Kepler-68d  &    580.00 $\pm$ 15.00 & 1.40 $\pm$ 0.03 &  0.18 $\pm$ 0.05 & 0.95 $\pm$  0.04 &  $M\sin i$ &  0.96 $\pm$ 0.27 & \citet{gil13} \\
Kepler-93c  &   1460.00             &  --            &        --       & 3.00             &  $M$     &  2.98 & \citet{mar}   \\ 
                \hline
        \end{tabular}
\end{table*}

\section{Conclusions}
\label{sec:conc}

Asteroseismology has recently detected oscillation frequencies of many host stars.
In this study, interior models with {\small {MESA}} code for 20 planet and planet-candidate solar-like oscillating host stars are constructed under
influence of these observational constraints.
Mass, radius, initial helium abundance, and  age of the host stars on the different evolutionary phases 
are derived from the constructed models. 
We also examine oscillation frequencies of these host stars observed by {\it {CoRoT}} and {\it {Kepler}}. Mean large and small separations between oscillation 
frequencies, and frequencies of the maximum amplitude are computed and used as constraints for the interior models. 
The reference frequencies also put very important constraints into interior models. 

We find that model mass range of the host stars is 0.74-1.55 $\rm M_{\sun}$. Among the host stars, KIC 6278762 and KIC 10666592 have 
the lowest (0.74 $\rm M_{\sun}$) and highest masses (1.55 $\rm M_{\sun}$), respectively.
KIC 6278762 also has the lowest model radius (0.75 $\rm R_{\sun}$) and is the oldest (11.7 Gyr) star among the host stars. 
Effective temperature and initial metallicity range of the host stars are
5000-6350 K and $Z_{0{\rm mod}}=0.012-0.026$, respectively. 

Most of the host stars have two reference minima, 
especially ${\nu_{\rm min0}}$ and ${\nu_{\rm min1}}$, in ${\Delta{\nu}_{\rm obs}}$ versus ${\nu_{\rm obs}}$ graph. 
Besides these minima, KIC~3632418, KIC~8866102, KIC~9592705, KIC~10666592, and KIC~11807274 have either entire or part of the  min2 glitch.
From the models with arbitrary mass and abundance, we confirm that ${\nu_{\rm min2}}$ is deeper than ${\nu_{\rm min0}}$ and ${\nu_{\rm min1}}$.
${{\rm min0}}$ of the model oscillation frequencies is very shallow in comparison to min1.  
In addition, high-frequency region in ${\Delta{\nu}}$ versus ${\nu}$ graph for the observed oscillation frequencies is fluctuated. 
Therefore, ${\nu_{\rm min0}}$ is either less or not confidential in some cases. 
Agreement between patterns of observed and model oscillation frequencies in ${\Delta{\nu}}$ versus ${\nu}$ graph,
in particular for the range around ${\nu_{\rm min1}}$  and ${\nu_{\rm min2}}$, reveals the appropriate model parameters.

We also compute fundamental properties of 34 planets and also four planet candidates. Orbital and fundamental parameters of the transiting planets and 
planet candidates are revised. 
Radius range of the transiting planets is 0.35-16.50 $\rm R_{\earth}$. Orbital inclination of the planets is approximately 90{\degr}. 
Semimajor axis range of the planets is
0.04-0.35 au. According to the values of the semimajor axis and radius, Kepler-2b is classified as hot Jupiter. We also derive fundamental parameters of the planet candidates.  
Their radius range is 0.55-3.15 $\rm R_{\earth}$.
While the maximum difference between estimated and transit radii is
about 25 per cent for the five planets in Kepler-444. 

Mass range of the planets is 0.29-3 $\rm M_{\rm jup}$. 
The estimated planetary mass of HD~52265b is 16 per cent higher than mass from the literature.

\begin{figure}
\includegraphics[width=\columnwidth]{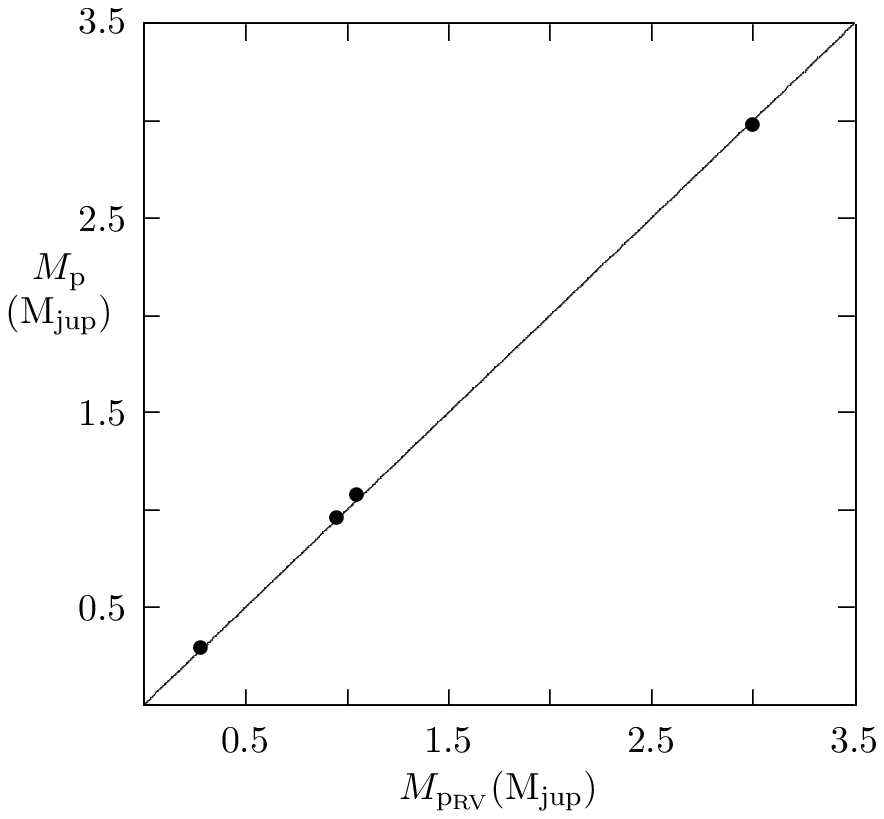}
\caption{Estimated planetary mass is plotted with respect to mass from RV data in $\rm M_{\rm jup}$ units.}
\label{fig:Mplitrv}
\end{figure}

\section*{Acknowledgements}
{The anonymous referee is acknowledged for his/her
suggestions which improved the presentation of the manuscript.}
We would like to thank Kelly Spencer for her kind help in checking the language of the revised manuscript.





\bsp	
\label{lastpage}

\begin{thebibliography}{99}

\bibitem[\protect\citeauthoryear{Angulo et al.}{1999}]{ang} Angulo C. et al., 1999, Nucl. Phys. A, 656, 3
\bibitem[\protect\citeauthoryear{Appourchaux et al.}{2012}]{app} Appourchaux T. et al., 2012, A\&A, 543, A54
\bibitem[\protect\citeauthoryear{Baglin et al.}{2006}]{bag} Baglin A., Michel E., Auvergne M., COROT Team, 2006, in Fletcher K., ed., Proc. SOHO 18/GONG 2006/HELAS I (ESA SP-624), Beyond the Spherical Sun. ESA, Noordwijk, p. 34
\bibitem[\protect\citeauthoryear{Ballard et al.}{2014}]{bal} Ballard S. et al., 2014, ApJ, 790, 12
\bibitem[\protect\citeauthoryear{Ballot et al.}{2011}]{bal11} Ballot J. et al., 2011, A\&A, 530, A97
\bibitem[\protect\citeauthoryear{Barclay et al.}{2015}]{barc} Barclay T. et al., 2015, ApJ, 800, 46
\bibitem[\protect\citeauthoryear{Bellinger}{2019}]{Bellinger} Bellinger E. P., 2019, MNRAS, 486, 4612
\bibitem[\protect\citeauthoryear{Benomar et al.}{2014}]{ben} Benomar, O., Masuda K., Shibahashi H., Suto Y., 2014, PASJ, 66, 94
\bibitem[\protect\citeauthoryear{B\"{o}hm-Vitense}{1958}]{bomv} B\"{o}hm-Vitense E., 1958, Zs. Ap., 46, 108
\bibitem[\protect\citeauthoryear{Borucki et al.}{2010}]{bor10} Borucki W. et al., 2010, Science, 327, 977
\bibitem[\protect\citeauthoryear{Brown et al.}{1991}]{bro} Brown T. M., Gilliland R. L., Noyes R. W., Ramsey L. W., 1991, ApJ, 368, 599
\bibitem[\protect\citeauthoryear{Bruntt et al.}{2012}]{bru} Bruntt H. et al., 2012, MNRAS, 423, 122
\bibitem[\protect\citeauthoryear{Campante et al.}{2015}]{cam} Campante T. L. et al., 2015, ApJ, 799, 170
\bibitem[\protect\citeauthoryear{Carter et al.}{2012}]{car} Carter J. A. et al., 2012, Science, 337, 556
\bibitem[\protect\citeauthoryear{Chaplin et al.}{2013}]{cha13} Chaplin W. J. et al., 2013, ApJ, 766, 101
\bibitem[\protect\citeauthoryear{Chaplin et al.}{2014}]{cha14} Chaplin W. J. et al., 2014, ApJS, 210, 1
\bibitem[\protect\citeauthoryear{Christensen-Dalsgaard}{2008}]{chris} Christensen-Dalsgaard J., 2008, Ap\&SS, 316, 113
\bibitem[\protect\citeauthoryear{Cyburt et al.}{2010}]{cyb} Cyburt R. H. et al., 2010, ApJS, 189, 240
\bibitem[\protect\citeauthoryear{Davies et al.}{2016}]{dav} Davies G. R. et al., 2016, MNRAS, 456, 2183
\bibitem[\protect\citeauthoryear{Deheuvels et al.}{2016}]{deh} Deheuvels S., Brand{\~{a}}o I., Silva Aguirre V., Ballot J., Michel E., Cunha M. S., Lebreton Y., Appourchaux T., 2016, A\&A, 589, A93
\bibitem[\protect\citeauthoryear{Escobar et al.}{2012}]{esc} Escobar M. E. et al., 2012, A\&A, 543, A96
\bibitem[\protect\citeauthoryear{Everett et al.}{2013}]{eve} Everett M. E. et al., 2013, ApJ, 771, 107
\bibitem[\protect\citeauthoryear{Gilliland et al.}{2013}]{gil13} Gilliland R. L. et al., 2013, ApJ, 766, 40
\bibitem[\protect\citeauthoryear{Huber et al.}{2013}]{hub} Huber D. et al., 2013, ApJ, 767, 127
\bibitem[\protect\citeauthoryear{Huber et al.}{2014}]{hub14} Huber D. et al., 2014, ApJS, 211, 2
\bibitem[\protect\citeauthoryear{Iglesias \& Rogers}{1993}]{igl93} Iglesias C. A., Rogers, F. J., 1993, ApJ, 412, 752
\bibitem[\protect\citeauthoryear{Iglesias \& Rogers}{1996}]{igl96} Iglesias C. A., Rogers, F. J., 1996, ApJ, 464, 943
\bibitem[\protect\citeauthoryear{Kjeldsen \& Bedding}{1995}]{kje.bed} Kjeldsen H., Bedding T. R., 1995, A\&A, 293, 87
\bibitem[\protect\citeauthoryear{Kjeldsen, Bedding \& Christensen-Dalsgaard}{2008}]{kje} Kjeldsen H., Bedding T. R., Christensen-Dalsgaard J., 2008, ApJ, 683, L175
\bibitem[\protect\citeauthoryear{Koch et al.}{2010}]{koc} Koch D. G. et al., 2010, ApJL, 713, L79
\bibitem[\protect\citeauthoryear{Kunz et al.}{2002}]{kun} Kunz R., Fey M., Jaeger M., Mayer A., Hammer J. W., Staudt G., Harissopulos S., Paradellis T., 2002, ApJ, 567, 643
\bibitem[\protect\citeauthoryear{Lebreton \& Goupil}{2014}]{lebgou14} Lebreton Y., Goupil M. J., 2014, A\&A, 569, A21
\bibitem[\protect\citeauthoryear{Lejeune, Cuisinier \& Buser}{1998}]{lej} Lejeune T., Cuisinier, F. Buser, R., 1998, A\&AS, 130, 65
\bibitem[\protect\citeauthoryear{Marcy et al.}{2014}]{mar} Marcy G. W. et al., 2014, ApJS, 210, 20
\bibitem[\protect\citeauthoryear{Mathur et al.}{2012}]{mat} Mathur S. et al., 2012, ApJ, 749, 152
\bibitem[\protect\citeauthoryear{Metcalfe et al.}{2014}]{met14} Metcalfe T. S. et al., 2014, ApJS, 214, 27
\bibitem[\protect\citeauthoryear{Molenda-{\.{Z}}akowicz et al.}{2013}]{mol} Molenda{-\.{Z}}akowicz J. et al., 2013, MNRAS, 434, 1422
\bibitem[\protect\citeauthoryear{Naef et al.}{2001}]{nae} Naef D., Mayor M., Pepe F., Queloz D., Santos N. C., Udry S., Burnet M., 2001, A\&A, 375, 205
\bibitem[\protect\citeauthoryear{Narita et al.}{2010}]{nar10} Narita N. et al., 2010, PASJ, 62, 779
\bibitem[\protect\citeauthoryear{Nutzman et al.}{2011}]{nut} Nutzman P. et al., 2011, ApJ, 726, 3  
\bibitem[\protect\citeauthoryear{P{\'{a}}l et al.}{2008}]{pal} P{\'{a}}l A. et al., 2008, ApJ, 680, 1450
\bibitem[\protect\citeauthoryear{Paxton et al.}{2011}]{pax11} Paxton B. et al., 2011, ApJS, 192, 39
\bibitem[\protect\citeauthoryear{Paxton et al.}{2013}]{pax13} Paxton B. et al., 2013, ApJS, 208, 49
\bibitem[\protect\citeauthoryear{Rowe et al.}{2014}]{row} Rowe J. F. et al., 2014, ApJ, 784, 45
\bibitem[\protect\citeauthoryear{Rowe et al.}{2015}]{row15} Rowe J. F. et al., 2015, ApJS, 217, 16
\bibitem[\protect\citeauthoryear{Santos et al.}{2013}]{san} Santos N. C. et al., 2013, A\&A, 556, A150
\bibitem[\protect\citeauthoryear{Seager \& Mall{\'{e}}n-Ornelas}{2003}]{sea} Seager S., Mall{\'{e}}n-Ornelas G., 2003, ApJ, 585, 1038
\bibitem[\protect\citeauthoryear{Silva Aguirre et al.}{2015}]{silva} Silva Aguirre V. et al., 2015, MNRAS, 452, 2127
\bibitem[\protect\citeauthoryear{Soriano et al.}{2007}]{sor} Soriano M., Vauclair S., Vauclair G., Laymand M., 2007, A\&A, 471, 885
\bibitem[\protect\citeauthoryear{Steffen et al.}{2012}]{ste12} Steffen J. H. et al., 2012, MNRAS, 421, 2342
\bibitem[\protect\citeauthoryear{Sullivan et al.}{2015}]{sul} Sullivan P. W. et al., 2015, ApJ, 809, 77
\bibitem[\protect\citeauthoryear{Ulrich}{1986}]{ulr} Ulrich R. K., 1986, ApJL, 306, L37
\bibitem[\protect\citeauthoryear{Van Eylen et al.}{2014}]{van} Van Eylen V. et al., 2014, ApJ, 782, 14
\bibitem[\protect\citeauthoryear{White et al}{2011}]{Whiteetal} White T. R. et al., 2011, ApJL, 742, L3
\bibitem[\protect\citeauthoryear{Wright et al.}{2011}]{wri} Wright D. J. et al., 2011, ApJ, 728, L20
\bibitem[\protect\citeauthoryear{Xie}{2014}]{xie} Xie J-W., 2014, ApJS, 210, 25 
\bibitem[\protect\citeauthoryear{Y{\i}ld{\i}z}{2015}]{yi15} Y{\i}ld{\i}z M., 2015, Res. Astron. Astrophys., 15, 2244
\bibitem[\protect\citeauthoryear{Y{\i}ld{\i}z et al.}{2014}]{yil14} Y{\i}ld{\i}z M., \c{C}elik Orhan Z., Aksoy \c{C}., Ok S., 2014, MNRAS, 441, 2148 (Paper I)
\bibitem[\protect\citeauthoryear{Y{\i}ld{\i}z, \c{C}elik Orhan \& Kayhan}{2015}]{yil15} Y{\i}ld{\i}z M., \c{C}elik Orhan Z., Kayhan C., 2015, MNRAS, 448, 3689 (Paper II)
\bibitem[\protect\citeauthoryear{Y{\i}ld{\i}z, \c{C}elik Orhan \& Kayhan}{2016}]{yil16} Y{\i}ld{\i}z M., \c{C}elik Orhan Z., Kayhan C., 2016, MNRAS, 462, 1577 (Paper III)


\end{thebibliography}
\end{document}